\documentstyle[12pt]{article}
\title{Higgs Model without Elementary Scalars}
\author{B.A. Arbuzov\\
{\it Skobeltsyn Institute for Nuclear Physics, Moscow State 
University}\\ {\it 119899 Moscow, Russia}}
\date{}
\newcommand{\be}{\begin{equation}}
\newcommand{\ee}{\end{equation}}
\newcommand{\beq}{\begin{eqnarray}}
\newcommand{\eeq}{\end{eqnarray}}
\newcommand{\nn}{\nonumber}
\newcommand{\bi}{\bibitem}
\begin{document}
\maketitle
\begin{quote}
A model of interaction of massless vector and spinor fields is 
considered. With the use of Bogolyubov quasi-average method we 
study a possibility of a dynamical breaking of the initial symmetry. 
Assuming the existence of effective cut-off $\Lambda$, 
we show, that there exists a solution, which breaks  
gauge symmetry of the theory. Instead of Higgs 
scalars fermion-antifermion tachion bound states are present here. 
As a result we have a theory with a massive vector field, a massive 
spinor and a composite scalar.
\end{quote}

The widely popular Higgs mechanism of the electroweak symmetry breaking 
needs initial scalar fields, which look rather less attractive, than 
the well-known gauge interactions of vector and spinor fields. 
Experimental facilities now approach the region of possible discovery of 
the Higgs. In view of these considerations it may be useful to 
study once more possibilities, which differ from the standard Higgs 
mechanism.

In the present note\footnote[1]{The work is supported in part by 
RFBR grant 98-02-16690 and grant "Universities of Russia" 990588.}
we consider a model, which might serve a substitute 
for the famous simple Higgs model~\cite{Higgs}. So, we consider $U(1)$ 
massless gauge field $A_\mu$ and also massless spinor field $\psi$, 
which interact in the followig way
\beq
& &L\,=\,\frac{\imath}{2}\bigl(\bar \psi \gamma_\rho \partial_\rho \psi\,- 
\,\partial_\rho \bar \psi \gamma_\rho \psi \bigr)\,-\,\frac{1}{4}
A_{\mu \nu}A_{\mu \nu}\,+\nn\\
& &+\,e_{L}\bar \psi_L \gamma_\rho \psi_L A_\rho\,
+\,\,e_{R}\bar \psi_R \gamma_\rho \psi_R A_\rho\,;\label{init}\\
& &A_{\mu\nu}\,=\,\partial_\mu A_\nu\,-\,\partial_\nu A_\mu\,;\nn
\eeq
where as usually
$$
\psi_L\,=\,\frac{1+\gamma_5}{2}\,\psi\,;\qquad 
\psi_R\,=\,\frac{1-\gamma_5}{2}\,\psi\,.
$$
Of course, the interesting possibility is the symmetric one
\be
e_{L}\,=\,e_{R}\,=\,e\,;\label{ee}
\ee
but we here take the general case and will not discuss the problem of 
the triangle axial anomaly, bearing in mind that in a 
more realistic model one always can arrange cancellation of the 
anomalies by a suitable choice of fermions' charges, in the same way 
as it occurs in the Standard Model.

Now we start to apply Bogolyubov quasi-average method~\cite{Bog}. In view 
of looking for symmetry breaking we add to~(\ref{init}) additional term
\be
\epsilon \cdot \bar \psi_L \psi_R\,\bar \psi_R \psi_L\,.\label{eRL}
\ee

Now let us consider the theory with $\epsilon \neq 0$, calculate necessary 
quantities (averages) and only at this stage take limit $\epsilon \to 0$. 
In this limit, according~\cite{Bog}, we come to quasi-averages, which 
not always coincide with the corresponding averages, which one obtains 
directly from the initial Lagrangian~(\ref{init}).

Because of additional term~(\ref{eRL}) the following vertices inevitably 
appear
\be
x\cdot \bar \psi_L \gamma_\rho \psi_L\,\bar \psi_R \gamma_\rho \psi_R\,; \qquad
\frac{y}{2}\cdot \bar \psi_L \gamma_\rho \psi_L\,
\bar \psi_L \gamma_\rho \psi_L\,;\qquad
\frac{z}{2}\cdot \bar \psi_R \gamma_\rho \psi_R\,\bar \psi_R \gamma_\rho \psi_R\,.
\label{psi}
\ee
These vertices should have form-factors, which define 
effective cut-off $\Lambda$. 
The origin of the cut-off is connected with (quite possible) self-consistent 
solution of the corresponding dynamical equations. Examples of 
such equations shows, that there appear decreasing functions of momentum 
variables, e.g. $p^2$, 
of the form (see, e.g.~\cite{AF}) 
$$
(x\,p^2)^\gamma\,exp\,\Bigl( -\, const\,(x\,p^2)^\rho\Bigr)\,;\qquad
|p^2| \to \infty\,;
$$
where $\gamma,\,\rho$ are some numbers, usually fractional, and $x$ 
denote typical dimensional constant, appearing in the model (in our case 
it may be some combination of $x,\,y,\,z$~(\ref{psi})). So we would 
expect $\Lambda$ to be of the order of magnitude of $1/x,\,1/y,\,1/z$. 
In any case, in our model we use some fixed cut-off value $\Lambda$. 
The model works in the region of momentum variables much less,than the 
cut-off: $|p^2| \ll \Lambda^2$.

We consider compensation equations~\cite{Bog, Bog2} 
(in other words, gap equations) for $x,\,y,\,z$ in one-loop 
approximation and obtain following set of equations
\beq
& & x\,=\,-\,\frac{\epsilon}{2}\,+\,\frac{6\,e_L\,e_R\,x}{16\pi^2}\,
\ln\frac{\Lambda^2}{\bar m^2}
\,+\,\frac{\Lambda^2}{16 \pi^2}\Bigl(\,-\, 3\,x^2\,-\,2\,x\,(y + z)
\Bigr)\,;\nn\\
& & y\,=\,-\,\frac{6\,e_L^2\,y}{16 \pi^2}\,\ln\frac{\Lambda^2}{\bar m^2}
\,+\,
\frac{\Lambda^2}{16 \pi^2}\Bigl( -\,x^2\,\Bigr)\,;\label{set}\\
& & z\,=\,-\,\frac{6\,e_R^2\,z}{16 \pi^2}\,\ln\frac{\Lambda^2}{\bar m^2}
\,+\,
\frac{\Lambda^2}{16 \pi^2}\Bigl( -\,x^2\,\Bigr)\,;\nn
\eeq 
Vector boson exchange corrections are calculated in Landau gauge. 
Here $\bar m$ is the largest of two would-be  masses: that of 
the gauge boson $M$ and the spinor one $m$. 
As we shall see relation $M\,\ll\,m\,\ll \,\Lambda$ is natural in the 
model, so in the following we assume $\bar m = m$. 
Let $X,\,Y,\,Z$ be dimensionless variables 
\be
X\,=\,x\,\frac{\Lambda^2}{16 \pi^2}\,;\qquad
Y\,=\,y\,\frac{\Lambda^2}{16 \pi^2}\,;\qquad
Z\,=\,z\,\frac{\Lambda^2}{16 \pi^2}\,;\label{X}
\ee
We shall consider charges $e_i$ to be small enough, so let us first solve 
set~(\ref{set}) for $e_i = 0$. At this stage we also set $\epsilon \to 0$. 
There is, of course, trivial solution
$x\,=\,y\,=\,z\,=\,0\,$. 
In addition we have two nontrivial solutions. 
\beq
& &Z_1\,=Y_1\,=\,-\,1\,;\qquad X_1\,=\,1\,;\label{1} \\ 
& &Z_2\,=Y_2\,=\,-\,\frac{1}{16}\,;\qquad X_2\,=\,-\,\frac{1}{4}\,.\label{2}
\eeq

As we shall see further, just the second solution~(\ref{2}) 
will be the most interesting. In this case it is important to take into 
account $e^2$ terms in set~(\ref{set}). Considering these terms as 
small perturbations, we obtain for the second solution~(\ref{2})
\beq
& &X\,=\,X_2\,+\,\Delta X\,;\qquad
Y\,=\,Y_2\,+\,\Delta Y\,;\qquad
Z\,=\,Z_2\,+\,\Delta Z\,;\nn\\
& &\Delta X\,=\,-\,\frac{3}{40 \pi^2}\,\ln \frac{\Lambda^2}{m^2}
\biggl(\,\frac{e_L^2+e_R^2}{4}\,-\,
e_L\,e_R\biggr)\,\,.\label{delta}
\eeq
Expressions for $\Delta Y,\,\Delta Z$ will be of no use in our discussion. 

Now let us consider scalar bound states $(\bar \psi_L \psi_R,\,\bar \psi_R
\psi_L)$. Without $e^2$ corrections we have from Bethe-Salpeter 
equation in one-loop approximation
\beq
& &g\,=\,-\,4\,X\,F(\xi)\,g\,;\qquad \xi\,=\,\frac{k^2}{4 \Lambda^2}\,;
\qquad\mu\,=\,\frac{m^2}{\Lambda^2};\label{eq0}\\
& &F(\xi)\,=\,1\, -\, \frac{5}{2}\,\xi\, +\, 2\xi\,
\ln\frac{4 \xi}{1 + \xi}\,;\qquad \mu \ll \xi\,<\,1\,;\nn\\
& &F(\xi)\,=\,1\, +\, \frac{1}{3}\,\xi\, +\, 2\xi\,
\ln \mu\,+\,O(\xi\mu,\,\xi^2)\,;\qquad \xi\leq \mu \,.\nn
\eeq
where $g\,=\,const$ is just the Bethe-Salpeter wave function. 
Here $k^2$ is the scalar state Euclidean momentum squared, that is 
$k^2 > 0$ means tachion mass of the scalar. 
Function $F(\xi)$ decreases 
from the value $F(0) = 1$ with $\xi$ increasing. We see, that 
for solution~(\ref{2}) we have bound state with $k^2 = 0$ in full 
correspondence with Bogolyubov-Goldstone theorem~\cite{Bog2}, \cite{Gold}.
As for the first solution~(\ref{1}), there is
 no solution of Eq.~(\ref{eq0}) at all.  So in the present note we 
concentrate our attention on the solution~(\ref{2}). Note, that there is 
an additional argument in favour of solution~(\ref{2}). Namely, 
values $X$ and especially $Y,\,Z$ are small enough, so we may 
expect, that many-loop terms will not influence results strongly. 

Now let us take into account vector boson corrections. Equation for 
the bound state~(\ref{eq0}) is modified due to two sources. The first one 
corresponds to modified expression~(\ref{delta}). The second one consists 
in loop $e^2$ corrections to Eq.~(\ref{eq0}). In Landau gauge 
there are only two nonzero such one-loop diagrams: the triangle 
one and the self-energy of the scalar. 
Scalar $\bar \psi_R\,\psi_L$ has evidently charge $e_L-e_R$. 
Then we have
\be
g\,=\,-\,4\,\biggl(\,-\,\frac{1}{4}\,+\,\Delta X\biggr)\,F(\xi)\,g\,+\,
\biggl(\frac{3 e_L e_R}
{16 \pi^2}\,\ln \frac{\Lambda^2}{m^2}\,+\, \frac{3 (e_L - e_R)^2}
{16 \pi^2}\,\ln \frac{\Lambda^2}{m^2}\biggr)\,g\,.\label{eqe}
\ee
We see, that for small $e_L,\,e_R$ possible eigenvalues $\xi$ are 
also small. Then we have following condition for an eigenvalue
\beq
& &F(\xi)\,\biggl(1\,+\,\frac{3}{80 \pi^2}\,\ln \frac{\Lambda^2}{m^2}\,
f(e_L,\,e_R)\biggr)\,=\,1;\label{eigen}\\
& &f(e_L,\,e_R)\,=\,\Bigl( 7\,e_L^2\,+7\,e_R^2\,-\,13\,e_L\,e_R \Bigr).
\nn
\eeq
The first point to be checked is the consequence of the 
Bogolyubov-Goldstone 
theorem, that is the zero mass eigenstate for symmetric case~(\ref{ee}).
We would expect, that substitution of values~(\ref{ee}) 
into~(\ref{eigen}) gives cancellation of all $e^2$ terms and thus leads 
to zero-mass eigenstate. In fact, there are large terms with 
opposite signs, but we do not obtain full cancellation.
Of course we have no doubt in validity of the theorem. The additional 
terms, which reduce the corresponding coefficient afore $e^2$ to zero, 
are connected with many-loop diagrams (one vector boson loop and 
other with interactions~(\ref{psi})). We see, that some 7 --8 \% 
change in coefficients in $f(e_L,\,e_R)$ reduce it to square of 
$(e_L - e_R)$. So we assume, that accuracy of our simplified one-loop 
calculations correspond to these values and it is at least not more than 
10\%. We have already noted, that such accuracy is natural for 
solution~(\ref{2}). Thus for 
qualitative  discussion of the model we shall use the simple one-loop
approximation.

Let us consider again equation~(\ref{eigen}) for scalar bound state. 
We see, that there is tachion bound state in case $e^2$  
contribution being positive. 
Really, the eigenvalue condition for small $e_i^2$ reads
\be
k^2\,=\,m_0^2\,=\,\frac{3\,A}{160 \pi^2}\,f(e_L,\,e_R)\,\Lambda^2\,
.\label{m0}
\ee
Here we keep main logarithmic terms. 
Thus provided $f(e_L,\, e_R) > 0$ we have scalar complex tachion 
$\phi$ with negative mass squared $-\,m_0^2$.
 
We have the following vertices of interaction of $\phi$ with spinors
\be
g\,\Bigl(\bar \psi_R\,\psi_L\,\phi\,+
\,\bar \psi_L\,\psi_R\,\phi^*\Bigr)\,.\label{fint}
\ee
We normalize $g$ by demanding the charge of $\phi$ to be $e_L - e_R$ 
(again in one-loop approximation). This gives
\be
g^2\,=\,\frac{32\,\pi^2}{\ln\,(\Lambda^2/m^2)}\,.\label{c}
\ee

Then we calculate box diagram with four scalar legs. This gives us 
effective constant $\lambda$, which enters into additional term 
\be
\Delta\,L\,=\,-\,\lambda\,(\phi^*\,\phi)\,(\phi^*\,\phi)\,;
\qquad
\lambda\,=\,\frac{32\,\pi^2}{3\,\ln\,(\Lambda^2/m^2)}\,.\label{lam}
\ee
Now we come to the usual Higgs model~\cite{Higgs} with $m_0^2$~(\ref{m0}), 
$\lambda$~
(\ref{lam}) and $\phi$ charge $e_L\,-\,e_R$.

Thus from expressions~(\ref{m0}, \ref{lam}) we have usual 
vacuum average of $\sqrt{2}\,Re\,\phi\,=\,\eta$
\be
\eta^2\,=\,\frac{m_0^2}{\lambda}\,=\,\frac{9\,f(e_L,e_R)}{5120\,\pi^4}\,
\Lambda^2\,\ln\,\frac{\Lambda^2}{m^2}\,.\label{eta}
\ee

The vector boson mass duly arises and it reads as follows
\be
M^2\,=\,\frac{9\,(e_L - e_R)^2\,f(e_L,e_R)}{5120\,\pi^4}\,\Lambda^2\,
\ln\,\frac{\Lambda^2}{m^2}\,.\label{M}
\ee

Interaction~(\ref{fint}) leads to spinor mass $m$
\be
m\,=\,\frac{g\,\eta}{\sqrt{2}}\,;\quad
m^2\,=\,\frac{9\,f(e_L,e_R)}{320\,\pi^2}\,\Lambda^2\,.\label{m}
\ee

Thus, we obtain the result, that initially massless model of interaction 
of a spinor with a vector becomes after the symmetry breaking just a 
close analog of the Higgs model. We have now vector boson mass~(\ref{M}), 
spinor mass ~(\ref{m}) and a scalar bound state with mass $\sqrt{2}\,m_0$, 
\be
m^2_H\,=\,2\,m_0^2\,=\,\frac{3}{80 \pi^2}\,f(e_L,\,e_R)\,\Lambda^2\,.
\label{mH}
\ee
  
The result is expressed in terms of parameters $e_L,\,e_R$. If we 
take parameters
\be
\alpha_L\,=\,\frac{e_L^2}{4 \pi}\,;\qquad
\alpha_R\,=\,\frac{e_R^2}{4 \pi}\,;\label{alpha}
\ee
to be small enough, relation $M\,\ll\,m\,\ll \,\Lambda$ is justified. 
Note interesting relation between $m_H$ and $m$
\be
m_H\,=\,\frac{2}{\sqrt{3}}\,m\,.\label{mHm}
\ee

Due to relation~(\ref{m}) 
$$
\ln\frac{\Lambda^2}{m^2}\,=\,\ln\Biggl(\frac{320\,\pi^2}
{9\,f(e_L, e_R)}\Biggr)\,;
$$
so, all masses are proportional to cut-off $\Lambda$. In addition to 
relation~(\ref{mHm}) we have following relations, which for the sake of 
simplicity we present here for antisymmetric case $e_L\,=\,-\,e_R\,=
\,e\,,\;\alpha_L = \alpha_R = \alpha$
\be
\eta^2\,=\,\frac{m^2\,}{16 \pi^2}\,\ln\Biggl(\frac{80 \,\pi}
{243\,\alpha}\Biggr)\,;
\qquad M^2\,=\,\frac{\alpha\,m^2}{\pi}\,\ln\Biggl(\frac{80 \,\pi}
{243\,\alpha}\Biggr)\,.\label{ratio}
\ee

We may consider the energy density of the scalar field
\be
{\cal E}\,=\,-\,\frac{m_0^4}{4\,\lambda}\,=\,-\,\frac
{27\,\Lambda^4\,f^2(e_L,e_R)}{32\,\pi^2\,(160\,\pi^2)^2}\,\ln\,
\frac{\Lambda^2}{m^2}\,;\label{dens}
\ee
A minimum of this function could fix the stable variant of the model. 
These considerations may help in application of the present method to 
more realistic electroweak models.

We would formulate qualitative result of the work as follows: 
in the massless model with Lagrangian~(\ref{init}) with 
$e_L \neq e_R$ there arises fermion-antifermion condensate, which 
defines masses $M,\,m,\,m_H$ according to~(\ref{mHm}, \ref{ratio}). 

Note, that variants of dynamical breaking of the electroweak symmetry 
without elementary scalars were considered in various aspects 
(see, e.g. paper~\cite{Arb}). The possibility  
scalars being composed of fundamental spinors was considered e.g. in 
well-known paper~\cite{Ter}.
  
The author is deeply grateful to R.N. Faustov, M.Z. Iofa and I.P. Volobuev 
for valuable discussions.

\end{document}